\documentclass[preprint,preprintnumbers,amsmath,amssymb,prb]{revtex4}

\usepackage{graphicx}
\usepackage{dcolumn}
\usepackage{bm}

\begin{document}
\title{A Finite-Size Scaling Study of a Model of Globular Proteins}
\author{D. L. Pagan
      \\  \emph{\small{ Department of Physics, Lehigh University, Bethlehem, P.A. 18015}}
      \\M. E. Gracheva
      \\ \emph{\small{Department of Mathematics, University of Minnesota, M.N. 55455}}
       \\  J. D. Gunton
      \\    \emph{\small{Department of Physics, Lehigh University, Bethlehem, P.A.
       18015}}}
\date{\today}

\begin{abstract}
 \noindent Grand canonical Monte Carlo
simulations are used to explore the metastable fluid-fluid
coexistence curve of the modified Lennard-Jones model of globular
proteins of ten Wolde and Frenkel (Science, \textbf{277}, 1975
(1997)). Using both mixed-field finite-size scaling and histogram
reweighting methods, the joint distribution of density and energy
fluctuations is analyzed at coexistence to accurately determine
the critical-point parameters. The subcritical coexistence region
is explored using the recently developed hyper-parallel tempering
Monte Carlo simulation method along with histogram reweighting to
obtain the density distributions. The phase diagram for the
metastable fluid-fluid coexistence curve is calculated in close
proximity to the critical point, a region previously unattained by
simulation.
\end{abstract}

\maketitle

\section{Introduction}

    \indent
     \indent Many pathological diseases, including sickle cell anemia~\cite{kn:Galkin} and genetic
     cataracts~\cite{kn:cat},
    are known to be caused by the crystallization of
    certain globular proteins. The importance of proteins is further exemplified by
    recent advances in genome sequencing, revealing that as much as
    ninety-eight percent
          of DNA may be involved in the regulation of their production.
            Exploring
    protein structure and activities (proteomics) is a
    growing research field and should help in our understanding of
    health and disease on a molecular basis~\cite{kn:web}.
    Advances in  decoding genomes, however, have far and away
    surpassed those in the determination of protein structure.
    The growth of high quality protein crystals from solution is important in determining structure
    and is known to depend sensitively on the initial conditions of the solution.
    Unfortunately,
    knowledge of the initial conditions necessary for optimal crystallization
    has not come easily~\cite{kn:mcpherson}.

  \indent Significant progress in understanding the relationship of the initial conditions
   to the crystal nucleation rates for globular proteins has been made in recent years. It
   was demonstrated in the theoretical work by Gast, Hall, and Russel~\cite{kn:gast} that the characteristics
   of the phase diagram of colloids depend sensitively on the range of attraction between
   the colloidal particles. It was
   found that for a colloid-colloid attractive interaction of very short range (less than
   some thirty percent of the colloidal diameter), there are  stable fluid and crystal phases
   and a metastable fluid-fluid phase located below the liquidus line. Other studies~\cite{kn:other}
   have found
   similar results. This has also been demonstrated
   in both experiment~\cite{kn:ilett} and simulation~\cite{kn:hagen}. Rosenbaum, Zamora,
   and Zukoski~\cite{kn:rosenbaum} linked the experimental observations of George and
   Wilson~\cite{kn:wilson} with those of colloids. They found that the narrow range in the second
   virial coefficient for which globular proteins crystallize map onto an effective temperature range
   for colloidal systems. The phase diagrams of the two systems were analogous.  ten Wolde and
   Frenkel~\cite{kn:tenwolde} then calculated the phase diagram and nucleation rate for a
   modified Lennard-Jones model of globular proteins, whose range of attractive interaction
   was small compared with the protein diameter.  In this seminal work they showed that the nucleation rate
   increased by many orders of magnitude in the vicinity of the critical point,
   suggesting a direct
   route to effective crystallization. Therefore, accurate
   knowledge of the region around the critical point provides important information
   regarding crystallization.

   \indent  ten Wolde and Frankel ~\cite{kn:tenwolde} studied a
    modified Lennard-Jones (MLJ) pairwise interaction model given
    by
    \newcolumntype{L}{>{$}1<{$}}
    \newenvironment{Cases}{\begin{array}{c}\{{1L}.}{\end{array}}
    \begin{equation}
               \large{V(r)} = \left\{ \begin{array}{cl}
                        \infty, & \mbox{$r < \sigma$}  \\
                        \frac{4\epsilon}{\alpha^{2}}(\frac{1}{[(r/\sigma)^{2}-1]^{6}}
                   -
                   \frac{\alpha}{[(r/\sigma)^{2}-1]^{3}})&\mbox{$r\ge\sigma$}.
                   \end{array} \right.
    \end{equation}

    \noindent V(r) is shown in Fig. 1, where \(\sigma \) denotes the hard-core diameter of the
    particles, \emph{r} is the interparticle distance, and
    \(\epsilon\) is the well depth. The width of the attractive
    well can be adjusted by varying the parameter \(\alpha\); it
    was tuned in such a way that the so-called stickiness parameter \(\tau\)~\cite{kn:rosenbaum} was
    equal to that produced by the hard-core Yukawa~\cite{kn:hagen} potential for
    \(k = 7\sigma^{-1} \)  at the metastable liquid-vapor
    critical point, where $k^{-1}$ is a measure of the range of the attractive part of the potential. In the Yukawa model, the phase diagram for \(k = 7\sigma^{-1}\)
    was found to be equivalent to that of globular proteins and
    maps
    onto those determined experimentally~\cite{kn:rosenbaum}. An advantage of the MLJ potential
    is that it lends itself naturally to both Monte Carlo and molecular dynamic simulations.
    Of particular interest both theoretically~\cite{kn:tenwolde,kn:oxtoby,kn:hagen} and
    experimentally~\cite{kn:exp1,kn:exp2,kn:exp3} is the metastable fluid-fluid curve of
    the phase diagram, for reasons noted above.
    ten Wolde and Frenkel~\cite{kn:tenwolde}
    determined the phase diagram for the above model using the
    Gibbs ensemble Monte Carlo (GEMC) method~\cite{kn:gibbs} where
    two coexisting phases are separated into two
    physically detached but thermodynamically connected boxes, the
    volumes of which are allowed to fluctuate under constant
    pressure. In the neighborhood of the critical point, however,
    GEMC cannot
    be relied upon to provide accurate estimates of the
    coexistence curve parameters~\cite{kn:reasons,kn:reasons2,kn:reasons3}. This is evident
     by considering
    the metastable region of the phase diagram~\cite{kn:tenwolde}.  The error bars in the data
    points grow larger as the critical point is approached; there
    are no
    data points in the immediate vicinity of the
    critical point. The purpose of this paper, then, is to
    accurately determine the critical point of the phase diagram
    of ten Wolde and Frenkel using finite size scaling techniques
    adapted for simple fluids by Bruce and Wilding and to accurately determine the corresponding
    subcritical region using the hyper-parallel tempering method.

     \begin{figure}
  \rotatebox{-90}{\scalebox{.35}{\includegraphics{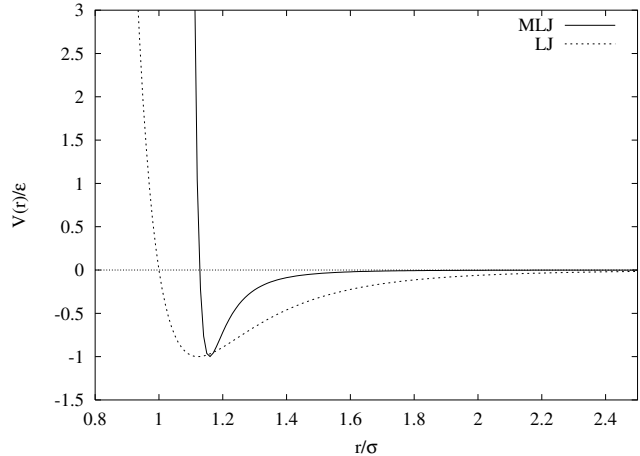}}}
  \caption{\label{fig:epsart}\small {The potential in Eq. (1) is shown as the  solid line. For comparison, the traditional Lennard-Jones potential is shown
  as the dashed line. The width of the MLJ was chosen so as to agree with the phase diagrams of experimentally determined
  globular proteins.}}
  \end{figure}

\section{Model}
\subsection{Theory}
    \indent
    \indent To study the critical region of the phase diagram, we use the
    Bruce and
   Wilding~\cite{kn:wilding} finite-size scaling method, along with
   histogram-reweighting~\cite{kn:swendsen} techniques, in conjunction with grand
   canonical Monte Carlo (GCMC) simulations~\cite{kn:liquids}. We
   write the reduced chemical potential as $\hat\mu =
   \mu/(k_{B}T)$ and the reduced well depth  as $\omega = \epsilon/(k_{B}T)$,
   with \emph{T} the temperature of the system. In what follows,
   we denote $\hat\mu$ as $\mu$ for simplicity.
   The critical point is
   identified by the critical values of the reduced chemical potential
   $\mu_{c}$ and well depth \(\omega_{c}\).
    The relevant scaling fields
   comprise linear combinations of \(\mu\) and \(\omega\):
   \begin{equation}
   \tau = \omega_{c} - \omega + s(\mu - \mu_{c})
   \end{equation}
   \begin {equation}
   h = \mu - \mu_{c} + r(\omega_{c} - \omega),
   \end{equation}

   \noindent with \(\tau\) and \emph{h} the thermal and ordering scaling
   fields, respectively. The parameters \emph{r} and \emph{s}
   are system-specific parameters, controlling the degree of \emph{mixing},
   and vanish identically when the Ising symmetry is present. Conjugate to
   these two fields are the scaling operators
    \begin{equation}
    \mathcal{M} = \frac{1}{1 - sr}\: [\rho - su]    \end{equation}
    \begin{equation}
    \mathcal{E} = \frac{1}{1 - sr}\: [u - r\rho],  \end{equation}

    \noindent with \(\mathcal{M} \) and \(\mathcal{E} \) the ordering and energy-like
    operators, respectively. The (dimensionless) number density and energy density are defined by
    \(\rho  = L^{-d}N \sigma^{d} \) and by \(u =
    U/(V\epsilon)\), respectively, with \emph{U} the total energy of the system,
    \(V = L^d\) \ the volume of the system, and \(\epsilon\) the
    well depth of the potential energy.

    We make the usual finite-size
    scaling \emph{ansatz}~\cite{kn:wilding}:

    \begin{equation}
    P_{L}(\mathcal{M},\mathcal{E}) \simeq\tilde{P}_{\mathcal{M},\mathcal{E}} (\Lambda_{\mathcal{M}}^\dag \; \delta{\mathcal{M}},\Lambda_{\mathcal{E}}^\dag \;
    \delta{\mathcal{E}}),
    \end{equation}

    \noindent where

    \begin{equation}
    \Lambda_{\mathcal{E}} = a_{\mathcal{E}}L^{1/\nu},\,
    \Lambda_{\mathcal{M}} = a_{\mathcal{M}}L^{d-\beta\nu},\,
    \Lambda_{\mathcal{M}}\Lambda_{\mathcal{M}}^\dag =\,
    \Lambda_{\mathcal{E}}\Lambda_{\mathcal{E}}^\dag\, = L^{d},
    \end{equation}

    \noindent and
   \begin{equation}
   \delta{\mathcal{M}} = \mathcal{M} - \langle\mathcal{M}\rangle_{c} \;\;  \delta{\mathcal{E}} = \mathcal{E} -
   \langle\mathcal{E}\rangle_{c},
   \end{equation}

   \noindent where the parameters \(a_{\mathcal{M}}\) and \(a_{\mathcal{E}}\) are
    non-universal scaling factors, $\beta$ and $\nu$ are the critical
    exponents
    for the coexistence curve and the correlation length~\cite{kn:stanley}, respectively, and
    the subscript \emph{c} in the above equations denotes that the averages are to be
    taken at criticality.
    From our
    simulations using GCMC, we obtain the joint probability
    distribution of \(\rho \) and \emph{u} at a particular point in
    parameter space of inverse temperature (\emph{T}) $\beta$ and reduced chemical potential $\mu$ from which we obtain
    the joint probability distribution of mixed operators via

    \begin{equation}
     P_{L}(\mathcal{M}, \mathcal{E})  = (1 - sr)\, P_{L}(\rho,u).
    \end{equation}

    \noindent Integrating out the energy-like dependence from the
    latter distribution gives
    \begin{equation}
    P_{L}(\mathcal{M}) = \int d\mathcal{E}\,P_{L}(\mathcal{M},\mathcal{E}).
    \end{equation}

    \noindent Using the fact ~\cite{kn:exper1,kn:exper2,kn:theory1,kn:theory2,kn:sim,kn:sim2}that the
    critical behavior of this
    distribution is in the same universality class as the Ising
    model, we match the above probability distribution to that of
    the universal fixed point function
    \begin{equation}
    \tilde{P}^{*}_{\mathcal{M}}(x) = \int \tilde{P}^{*}_{\mathcal{M},
    \mathcal{E}}(x,y)\,dy,
    \end{equation}
    \noindent which is known from an independent study~\cite{kn:lattice}. The
    critical point of the fluid can thus be estimated by tuning
    the temperature \emph{T}, chemical potential $\mu$, and field-mixing
    parameter \emph{s} such that \(P_{L}(\mathcal{M})\)
    collapses onto \(\tilde{P}^{*}_{\mathcal{M}}(x)\).

    To save computer-time and to facilitate the matching process,
    we employ histogram-reweighting~\cite{kn:swendsen} in lieu of
    performing tedious simulations.
    Once the joint probability distribution at a particular point in
   parameter space \((\beta,\mu)\) is obtained in one simulation run, information around its neighboring points in
   parameter space \((\beta',\mu')\) is extracted using

    \begin{eqnarray}
    &&P^{(\beta',\mu')}_{L}(\rho,u) = \nonumber \\
    && \frac{\exp[(\mu' - \mu)\rho V - (\beta' - \beta)u
    V]P^{(\beta, \mu)}_{L}(\rho,u)}{\sum \exp[(\mu' - \mu)\rho V - (\beta' - \beta)u
    V]P^{(\beta, \mu)}_{L}(\rho,u)}.
    \end{eqnarray}



    \noindent Histogram-reweighting provides an accurate
    determination of the new probability distribution and is well suited for Monte Carlo
    studies. Its limitations are discussed elsewhere~\cite{kn:limit}.

    In the subcritical region, we employ hyper-parallel
    tempering~\cite{kn:yan}
    Monte Carlo (HPTMC) to sample the joint probability distribution.
    Below the critical point, density fluctuations are no longer
    large enough for the joint probability distribution to be accessible by standard Monte Carlo
    simulations. Thus, a free energy barrier exists below $T_{c}$
    which needs to be overcome if one is to sample both coexisting phases. HPTMC allows one to effectively
    tunnel through this barrier by swapping particle
    configurations between different simulations (replicas) at
    different state points. Other techniques~\cite{kn:berg} sample
    this region by artificially lowering this barrier.

    In the grand canonical ensemble, the partition function can
    be written as

    \begin{equation}
       Z = \sum_{x} \Omega(x) exp[-\beta U(x) + \mu N(x)],
    \end{equation}

    \noindent where {x} denotes the state of the system, $\Omega (x)$ is the density of states, $U(x)$ is the
    total energy of state \emph{x}, and $N(x)$ is the number of particles in state \emph{x}, and all
    other variables are defined as above. In accord with practice~\cite{kn:yan}, we consider a composite ensemble
    consisting of \emph{M} non-interacting replicas, each at
    a different set of state points. The partition function of the
    composite ensemble is specified by

    \begin{equation}
        Z_c = \prod_{i=1}^{M} Z_{i},
    \end{equation}

    \noindent where the complete state of the composite ensemble is
    given by
    \begin{equation}
        \textbf{x} = (x_{1}, x_{2}, ... , x_{M}),
    \end{equation}

    \noindent with $x_{i}$ denoting the state of the $i^{th}$ replica. The unnormalized probability
    density of the composite state \textbf{x} is
    given by

    \begin{equation}
     p(\textbf{x}) = \prod_{i=1}^{M} exp[-\beta U(x_{i}) + \mu N(x_{i})].
    \end{equation}

    To sample configurations from the composite ensemble, a Markov
    chain is constructed to generate configurations according to
    the limiting function in Eq. (16). In the Markov chain, two types of
    trial moves are employed: 1) within each replica,
    insertion/deletion trial moves are attempted according to
    standard Monte Carlo as adapted for use in the grand canonical
    ensemble~\cite{kn:liquids}, and 2) Configuration swaps are attempted between
    pairs of replicas \emph{i} and \emph{i}+1 such that

    \begin{equation}
    x_{i}^{new} = x_{i + 1}^{old}, \;\;\; x_{i+1}^{new} = x_{i}^{old}.
    \end{equation}

    To enforce a detailed-balance condition, the pair of replicas
    that are attempted to be swapped are chosen at random, and the
    trial swap is accepted with the probability

    \begin{equation}
    p_{acc}(x_{i} \leftrightarrow x_{i+1}) = min[1, exp(\Delta \beta \Delta U - \Delta \mu \Delta
    N)],
    \end{equation}

    \noindent where $\Delta \beta = \beta_{i+1} - \beta_{i}$, $\Delta U = U(x_{i+1}) -
    U(x_{i})$, $\Delta \mu = \mu_{i+1} - \mu_{i}$, and $\Delta N = N(x_{i+1}) - N(x_{i})$.
    Once joint probability distributions are obtained in this way, histogram
    reweighting is applied to obtain coexistence according to Eq. (12) and an
    "equal-weight"~\cite{kn:hill} construction:
    \begin{equation}
    \int_{0}^{\langle\rho\rangle} P_{L}^{(\beta' \mu')}(\rho) = 0.5.
    \end{equation}

    \noindent It should be recognized that the average density $
    \langle\rho\rangle$ in the upper limit of the integral in
    Eq. (19) is itself a function of temperature and chemical
    potential and can therefore be obtained from the first moment
    of the reweighted histogram.

\subsection{Computational Details}
    \indent
    \indent We studied a system of N particles contained in a
    three-dimensional, periodic cubic simulation cell having a
    volume \(V = L^{d}\). Two particles separated by a distance
    \emph{r}
    interact via the modified Lennard-Jones (MLJ) pair potential
    given in Eq. (1), where \(\epsilon\) and \(\sigma\) denote the
    energy and length scales, respectively. The total energy, \emph{U},
    is obtained by summing over all distinct pairs of particles.
    We employ a truncated, unshifted version of Eq. (1) using a
    cutoff radius \(r_{c} = 2.0\,\sigma\), in accord~\cite{kn:foot} with ten Wolde and Frenkel. The simulations
    were performed on system sizes of \(L = 6\,\sigma, 7\,\sigma,
    8\,\sigma\), and \(10\,\sigma\), implemented
    using the Metropolis Monte Carlo method as adapted for use in
    the grand canonical ensemble~\cite{kn:liquids} with a constant volume $L^{d}$, inverse
    temperature (\emph{T}) \(\beta\), and reduced chemical potential
    \(\mu\). The thermodynamic potential needed is actually \(\mu^{*} =\mu - 3\;ln[\Lambda/\sigma]\)
    (where $\Lambda$ is the thermal deBroglie wavelength)~\cite{kn:liquids}. As in other implementations~\cite{kn:masha}, only particle insertion
    and deletion steps were employed, particle displacements being
    realized within the cell from a succession of such steps. In our simulations, equilibrium times used were approximately
    two million steps and production runs ranged from 500 million steps for the smaller system sizes to one billion
    steps for the higher system sizes. Such a high number of
    steps was needed in order to obtain smooth distributions at
    the low temperatures we are studying. We emphasize
    that the true equilibrium state of the system is that of fluid
    and solid coexistence. GCMC, however, is limited to 'low'
    densities and, thus, the solid region of the phase diagram is
    unaccessible.

    Using a previous estimate of the critical
    temperature~\cite{kn:thesis} for this model, we first attempted to
    locate the critical point by tuning the reduced chemical potential
    $\mu^{*}$ until the density distribution exhibited a double peaked
    structure. Once obtained, longer runs were performed to
    accumulate better statistics. To obtain two-phase coexistence and appropriate values
    of the field-mixing parameter \emph{s}, we adopted the criterion~\cite{kn:wilding}
    that the order distribution $P_{L}(\mathcal{M}) = \int d\mathcal{E}\, P_{L}(\mathcal{M},\mathcal{E})$
    be symmetric in $\mathcal{M\, -\, }\langle\mathcal{M}\rangle$. This criterion
    is the counterpart of the coexistence symmetry condition for the Ising model magnetization
    distribution. Having obtained in this manner
    a two-phase distribution near the critical point, we then
    matched the order-operator distribution $P_{L}(\mathcal{M})$ to
    that of the universal fixed point distribution $\tilde{P}^{*}_{\mathcal{M}}(x)$. Employing histogram-reweighting, we tuned the
    chemical potential $\mu^{*}$, temperature \emph{T}, and mixed-field
    parameter \emph{s} until our distribution $P_{L}(\mathcal{M})$
    collapsed onto that of the fixed point distribution. Once that was attained, we then attempted to match
    $P_{L}(\mathcal{E})$ to the corresponding energy-like fixed point distribution $\tilde{P}^{*}_{\mathcal{E}}(y)$
    by tuning the field-mixing parameter \emph{r}. This
    procedure was repeated for the various system sizes studied.

    In the subcritical region, we used six replicas for the $L =
    7\,\sigma$ system to obtain joint probability distributions. Small
    runs of approximately five million steps were first performed to
    optimize the choice of $\mu^{*}$ for each replica. At
    low temperatures, $\mu^{*}$ was chosen so that the
    density distribution was biased toward a high-density peak;
    conversely, at high temperatures, $\mu^{*}$ was chosen so that
    low-density peaks were favored. Such a choice of $\mu^{*}$ allowed
    high-density configurations to be "melted" when passed to high
    temperatures, allowing for a fuller exploration of
    configuration space~\cite{kn:yan}. Also, we chose these values so that a high
    swap frequency between replicas was realized. On average, replicas were swapped
    twenty-five percent
    of the time.  Once these criteria were satisfied, Monte Carlo runs
    were extended to approximately 150 million steps to acquire
    good statistics. Histogram-reweighting was then applied to the
    resulting joint probability distributions to obtain coexistence
    in the subcritical region in accordance with Eq. (12) and Eq. (19).

\section{Results}

    \indent
    \indent The resulting density distributions obtained in
    the prescribed manner at the size-dependent critical point $T_{c}(L)$ and
    $\mu^{*}_{c}(L)$ are shown in Fig. 2. As is evident, the distributions become
    narrower with increasing system size $L$, approaching the
    limiting form of the fixed point distribution
    $\tilde{P}^{*}_{\mathcal{M}}(x)$. The corresponding energy-density distributions are
    shown in Fig. 3.

    Both distributions show an asymmetry, due to field-mixing effects, with that of $P_{L}(u)$
    being much more pronounced. These effects die off with
    increasing L, so that the limiting forms of both $P_{L}(\rho)$ and
    $P_{L}(u)$ match the fixed point distribution
    $\tilde{P}^{*}_{\mathcal{M}}(x)$. As noted, this limiting form is easily
    recognizable for the density distributions. However, the
    limiting form of the energy density distributions does not
    follow this pattern. This difference is attributable to the
    coupling that occurs for asymmetric systems between the
    ordering operator and energy-like operator fluctuations. Those
    of $\mathcal{M}$ dominate at large L, while those in
    $\mathcal{E}$ do not~\cite{kn:wilding_fluctuate,kn:wilding_finite}. Thus, a
    'background' effect perturbs the energy-density distributions.

     \begin{figure}
     \rotatebox{-90}{\scalebox{.35}{\includegraphics{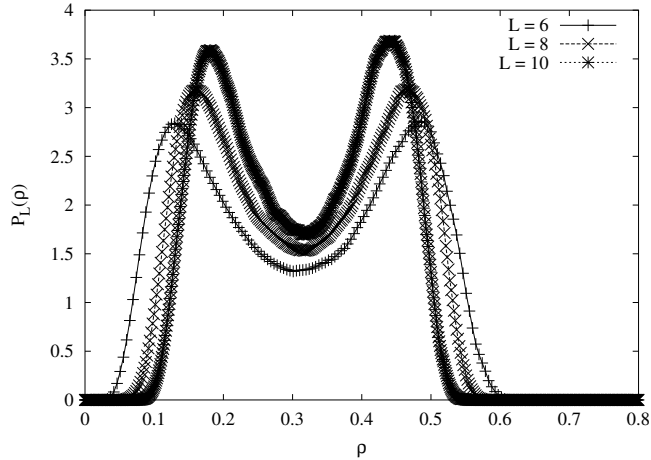}}}
     \caption{\label{fig:epsart}\small {Density distributions at $T_{c}(L)$ and $\mu^{*}_{c}(L)$ for the
     system sizes $L = 6\sigma, 8\sigma,$ and $10 \sigma$. For clarity, the distribution
     corresponding to the $L = 7$ system size is not shown.}}
     \end{figure}

    \begin{figure}
     \rotatebox{-90}{\scalebox{.35}{\includegraphics{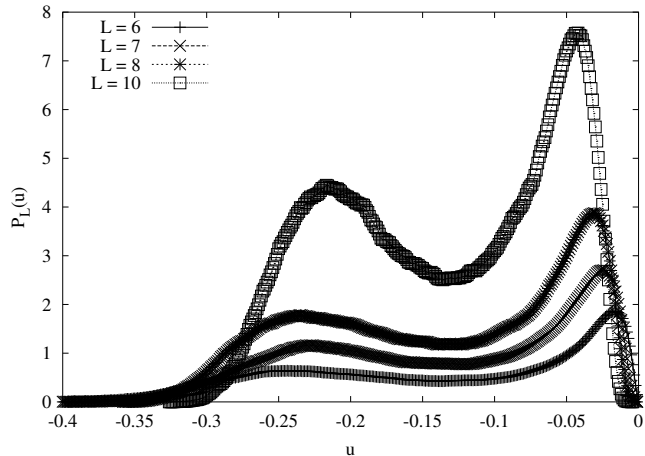}}}
     \caption{\label{fig:epsart}\small {Energy distributions at $T_{c}(L)$ and $\mu^{*}_{c}(L)$ for the
     system sizes $L = 6\sigma, 7\sigma, 8\sigma,$ and $10 \sigma$.}}
     \end{figure}

    Our estimates for the fixed point distribution are
    shown in Fig. 4. As is seen from the figure, the agreement is
    good for all system sizes studied. This matching alone allows
    us to accurately determine the critical point $T_{c}(L)$
    and $\mu^{*}_{c} (L)$ and to also obtain good estimates of
    the field-mixing parameter \emph{s}. Though matching of
    $P_{L}(\mathcal{E})$ to $\tilde{P}^{*}_{\mathcal{E}}(y)$
    should also give good estimates of the critical point, and the
    field-mixing parameter \emph{r}, fluctuations in the energy-like operator $\mathcal{E}$
    are relatively weak and therefore do not allow for good
    matching~\cite{kn:micro}. It is unknown presently how to remove
    this background effect for fluid-systems, though it has been
    removed for the Ising model~\cite{kn:micro}. Nevertheless, we
    can still obtain a rough estimate for the field-mixing
    parameter \emph{r} by observing that the distributions
    $P_{L}(\mathcal{E})$ should have a shape similar to
    $\tilde{P}^{*}_{\mathcal{E}}(y)$. For completeness, we include
    these curves along with the energy-fixed point function in
    Fig. 5.

    \begin{figure}
     \rotatebox{-90}{\scalebox{.35}{\includegraphics{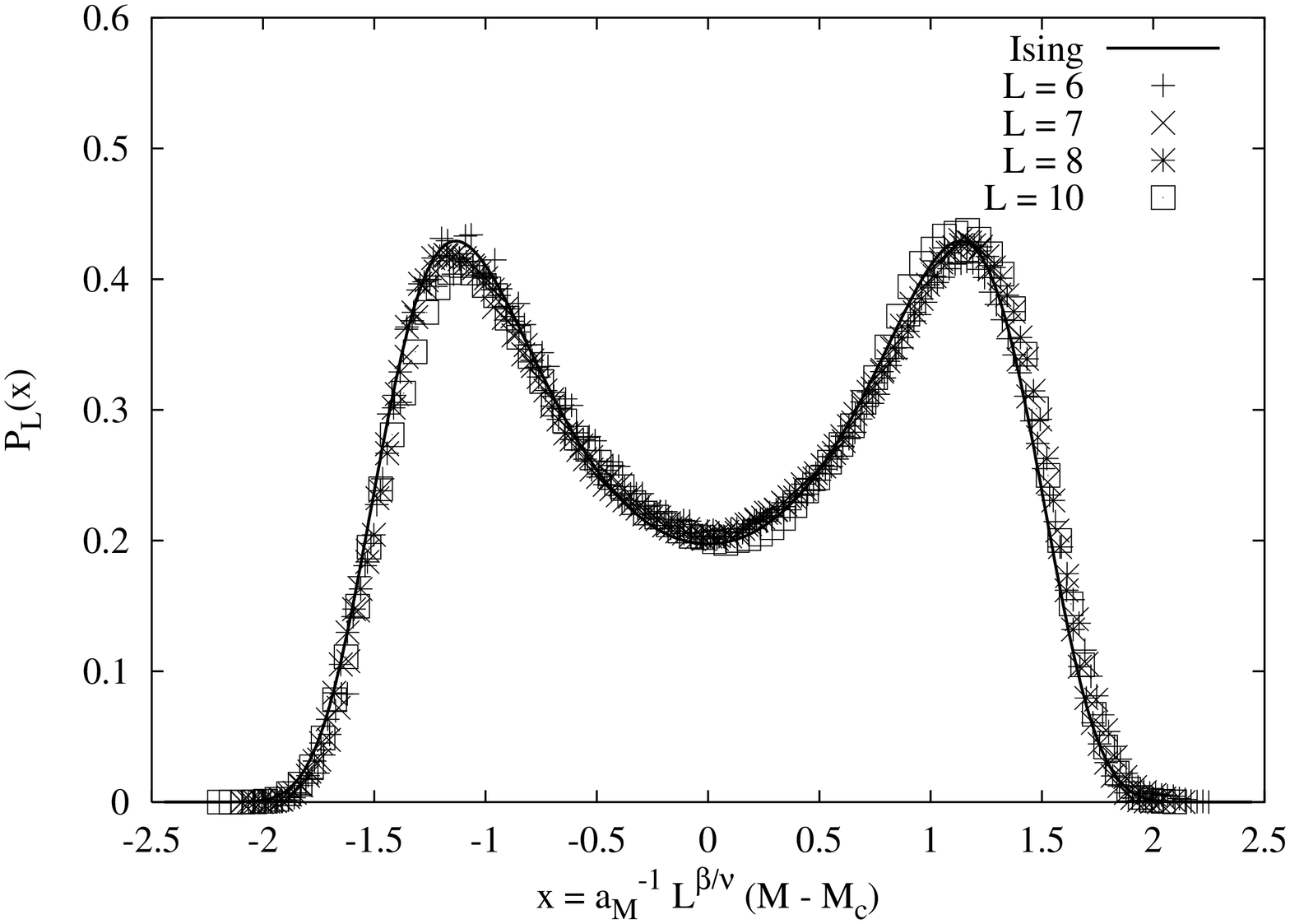}}}
     \caption{\label{fig:epsart}\small {The measured form of the ordering operator distribution
     $P_{L}(\mathcal{M})$ for the system sizes studied. Shown for comparison is the universal fixed-point ordering operator distribution $\tilde{P}^{*}_{\mathcal {M}}(x)$ (solid line).
     The data have been expressed in terms of the scaled variable $x = a_{\mathcal{M}}^{-1}L^{\beta/\nu} (\mathcal{M} - \mathcal{M}_{c} )$, where $a^{-1}_{\mathcal{M}}$ was chosen such that the
     distributions have unit variance.}}
     \end{figure}

    \begin{figure}
     \rotatebox{-90}{\scalebox{.35}{\includegraphics{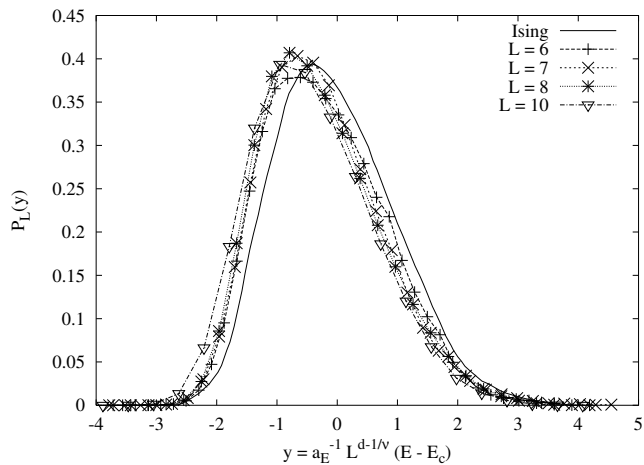}}}
     \caption{\label{fig:epsart}\small {The measured form of the energy-like operator distribution
     $P_{L}(\mathcal{E})$ for the system sizes studied. Shown for comparison is the universal fixed-point ordering operator distribution $\tilde{P}^{*}_{\mathcal {E}}(y)$ (solid line).
     The data have been expressed in terms of the scaled variable $y = a_{\mathcal{E}}^{-1}L^{d - 1/\nu} (\mathcal{E} - \mathcal{E}_{c} )$, where $a^{-1}_{\mathcal{E}}$ was chosen such that the
     distributions have unit variance. The lines are a guide to the eye.}}
     \end{figure}

    From estimates of the finite critical-point temperatures
    $T_{c}(L)$, we can estimate the critical-point for the infinite-volume system.
    Since contributions to $P_{L}(\mathcal{M})$ from finite values
    of $\tau$ grow with system size like $\tau L^{1/\nu}$, the
    matching condition results in a deviation from the true
    critical point $T_{c}(\infty)$ as
    \begin{equation}
        T_{c}(\infty) - T_{c}(L) \propto L^{-(\theta +
        1)/\nu},
    \end{equation}

    \noindent where $\theta$ is the universal correction to scaling exponent~\cite{kn:theta} and $\nu$ is
    the critical exponent for the correlation length~\cite{kn:stanley}. In Fig. 6, we plot the apparent critical temperature
    $T_{c}(L)$ as a function of $L^{-(\theta + 1)/\nu}$. By
    extrapolating to infinite volume, we arrive at an estimate for
    the true critical temperature. Similar arguments in accounting for field-mixing effects~\cite{kn:wilding_fluctuate} apply in obtaining the true critical density
    via
    \begin{equation}
        \langle\rho\rangle_c(L) - \langle\rho\rangle_{c}(\infty)
        \propto L^{-(d - 1/\nu)},
    \end{equation}
    where \emph{d} is the dimensionality of the system. In Fig. 7, $\langle\rho\rangle_{c}(L)$ is plotted as a
    function of $L^{-(d - 1/\nu)}$ to obtain an estimate of the
    true critical density. We summarize our findings in Table I.

     \begin{figure}
     \rotatebox{-90}{\scalebox{.35}{\includegraphics{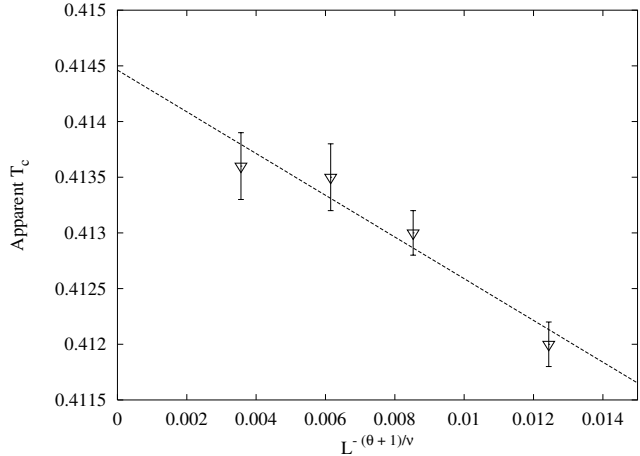}}}
     \caption{\label{fig:epsart}\small {The apparent reduced critical temperature plotted as a
     function of $L^{-(\theta +1)/\nu}$, with $\theta = 0.54~\cite{kn:theta}$ and $\nu =0.629$~\cite{kn:values}. The extrapolation
      to infinite volume yields the estimate $T_{c} = .4145(5)$.}}
     \end{figure}

    \begin{table}
    \caption{\label{tab:table1}Summary of results for location of the critical point}
    \begin{ruledtabular}
    \begin{tabular}{cccccc}
    L & $T_{c}(L)$ & $\mu^{*}_{c}(L)$ & $\langle\rho\rangle_{c}(L)$ & s & r \\
    \hline 6 & .4120$(3)$ & -2.939$(1)$ & .312$(2)$ & -0.120$(3)$ & -0.65$(5)$ \\
    7 & .4130$(2)$ & -2.932$(2)$ & .312$(2)$ & -0.115$(2)$ & -0.65$(5)$ \\
    8 & .4135$(3)$& -2.930$(2)$ & .318$(4)$ & -0.110$(1)$ & -0.65$(5)$ \\
    10 & .4136$(6)$ & -2.932$(3)$ & .315$(5)$ & -0.110$(1)$ &  -0.65$(5)$ \\
    \end{tabular}
    \end{ruledtabular}
    \end{table}

     \begin{figure}
     \rotatebox{-90}{\scalebox{.35}{\includegraphics{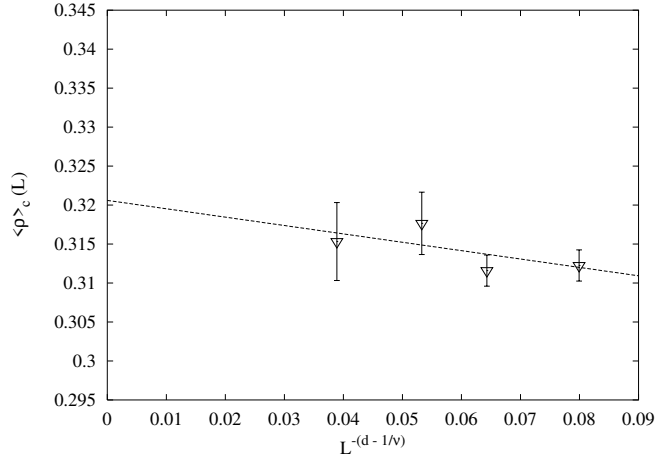}}}
     \caption{\label{fig:epsart}\small {The measured average density $\langle\rho\rangle_{c}(L)$ at the L-dependent critical point expressed as a function of $L^{-(d - 1)/\nu}$.
     Extrapolation to infinite volume yields the estimate $\rho_{c} = .3206(4)$.}}
     \end{figure}

     \begin{figure}
     \rotatebox{-90}{\scalebox{.35}{\includegraphics{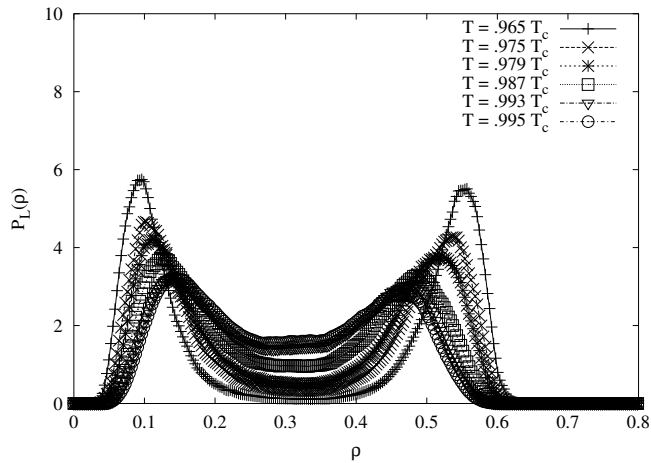}}}
     \caption{\label{fig:epsart}\small {The coexistence density distributions for
     the $L = 7\,\sigma$ system size for a range of subcritical temperatures, obtained using HPTMC as described in the text.}}
     \end{figure}

    \begin{table}
    \caption{\label{tab:table1}Values of T and $\mu^{*}$ used in HPTMC. Also
    shown are the reweighted chemical potentials $\mu^{*}_{R}$
    used in obtaining the coexisting densities $\rho_{v}$ and $\rho_{l}$.}
    \begin{ruledtabular}
    \begin{tabular}{cccccc}
    Replica & $T$ & $\mu^{*}$ & $\mu^{*}_{R}$ & $\rho_{v}$ & $\rho_{l}$\\\hline
    1 & .400 & -3.110 & -3.137 & .120 & .531\\
    2 & .404 & -3.040 & -3.071 & .140 & .510\\
    3 & .406 & -3.010 & -3.041 & .149 & .495\\
    4 & .409 & -2.970 & -2.995 & .170 & .473\\
    5 & .4115 & -2.947 & -2.957& .186 & .444\\
    6 & .4123 & -2.941 & -2.946& .192 & .441\\
    \end{tabular}
    \end{ruledtabular}
    \end{table}

   Using HPTMC, we were able to obtain joint distribution
   functions corresponding to points in the subcritical region of
   the phase diagram of ten Wolde and Frenkel. The choices we
   employed for the values of the temperature \emph{T} and
   chemical potential $\mu^{*}$ of each replica are displayed in
   Table II, along with the coexisting chemical potentials $\mu^{*}_{R}$
   and coexisting densities $\rho_{v}$ and $\rho_{l}$, obtained after reweighting. The corresponding density distributions
   in this region are
   shown in Fig. 8, where the temperatures are expressed in terms of the infinite-volume critical temperature.
   It can be seen that for temperatures even as
   high as $T = .965\,T_{c}$, the system has a vanishingly small
   probability of visiting a state between the two coexisting
   phases. This implies, as stated previously, that one cannot
   expect standard Monte Carlo simulations to obtain these
   subcritical distributions.

   In Fig. 9, we plot the temperatures
   used in Table II as a function of density, where we have calculated the average
   densities for each of the two phases from the distributions
   shown in Fig. 8. These new results are consistent with the earlier
   results for the metastable coexistence curve of
   the phase diagram~\cite{kn:tenwolde} and extend much closer to
   the critical point.  We also show our attempt to fit this subcritical density data
   to a power-law of the form $\rho_{\pm} - \rho_{c} = A \,|T - T_{c}| \:\pm\: B \,|T - T_{c}|^{\beta}$, where $T_{c}$ and $\rho_{c}$ are
   are our extrapolated values as $L \rightarrow \infty$ and $\beta = .3258~\cite{kn:values}$ is the Ising exponent.
   Although the data are in reasonable agreement with this fit,
 note
   that we have not accounted for possible finite-size effects, preventing this from being a
   definitive test.  Fig. 10 shows the
   corresponding chemical potential as a function of temperature,
   which obeys a linear relationship in this region.  Experimental~\cite{kn:schurt}
   investigations on the bovine lens protein, $\gamma_{II}-crystallin$, have
   been performed near the critical point. They report values for the
   critical isothermal compressibility $\gamma = 1.21 \pm 0.05$ and the critical correlation
   length $\nu = 0.68 \pm 0.1$. Both results are compatible
   to three-dimensional Ising model values.

    \begin{figure}
     \rotatebox{0}{\scalebox{.5}{\includegraphics{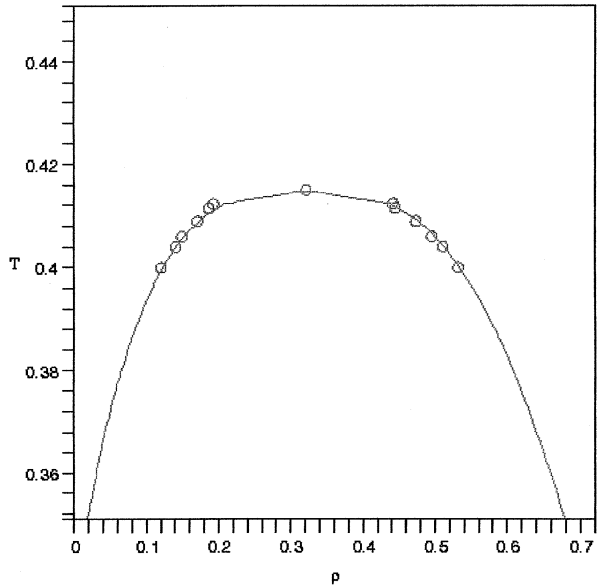}}}
     \caption{\label{fig:epsart}\small {Temperature plotted vs. density, as obtained from
     HPTMC simulations for the $L = 7\,\sigma$ system size shown with the critical point $T_{c}(\infty)$ and the corresponding critical
     density $\rho_{c}$ as obtained by Eqs. (20) and (21). Also shown (solid
     line) is a fit to the data, with $\beta = .3258~\cite{kn:values}$.
     }}
     \end{figure}

     \begin{figure}
     \rotatebox{-90}{\scalebox{.32}{\includegraphics{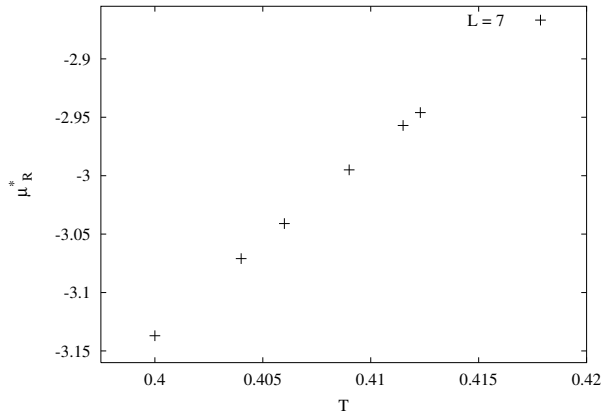}}}
     \caption{\label{fig:epsart}\small {The chemical potential $\mu^{*}_{R}$ vs. $ T $ for the system size $L = 7\,\sigma $ for
     the coexisting subcritical region obtained using HPTMC as
     described in the text. Errors do not exceed symbol size.
     }}
     \end{figure}

 \section{Conclusion}
    \indent
    \indent In this work, we have employed mixed-field finite-size scaling techniques and
    histogram extrapolation methods to obtain accurate
    estimates for the critical-point parameters of the truncated,
 unshifted MLJ fluid with $r_{c} = 2.0\,\sigma$. Our
    measurements allow us to pinpoint the critical-point
    parameters to within high accuracy. A previous estimate put the critical point at $T_{c} =
    .420$~\cite{kn:thesis}, slightly higher than our estimate of
    $T_{c} (\infty) = .4145(5)$. In the near subcritical region, we have explored the phase
    diagram and obtained data in a region where no prior estimates
    were available. HPTMC proved to be an efficient means toward
    this end.


   \begin{acknowledgements}
   This work has been supported by an NSF Grant, DMR-0302598. We would
   like to thank D. Frenkel and P. R. ten Wolde for helpful
   discussions concerning this project.
   \end{acknowledgements}


\begin{thebibliography}{99}
\bibitem{kn:foot}The potential was shifted in ref. [11].
\end{thebibliography}
\end{document}